\pdfoutput=1
\documentclass{article}

\usepackage{graphicx} 
\usepackage{placeins}
\begin{document}

\title{The PADME experiment}
\author{G. Piperno for the PADME Collaboration  \\
\footnotesize{\em INFN, Laboratori Nazionali di Frascati, Via E. Fermi, 40 -- I-00044 Frascati (Rome), Italy}
\\
}
\maketitle


\begin{abstract}
The PADME experiment, hosted at the Laboratori
Nazionali di Frascati, will search for a Dark Photon that decays
in invisible channels with a mass up to $23.7\,\mbox{MeV}$ and coupling
constant down to $10^{-3}$.

\end{abstract}

\section{Introduction}

Since first cosmological evidences, the the Dark
Matter (DM) direct detection continues to remain an open issue. This
puzzle can be solved hypothesizing that the DM does not directly interact
with the Standard Model (SM) gauge fields, but only by means of ``portals''
that connect our world with this dark sector. The simplest model adds
a U(1) symmetry and its corresponding vector boson $A'$ \cite{key-1}:
SM particles are neutral under this symmetry, while the new field
couple to SM with an effective charge $\varepsilon e$ and for this
reason is often called Dark Photon (DP).

Recently has been noted that an $A'$ with mass in the range $1\,\mbox{MeV}$ to $1\,\mbox{GeV}$
and constant $\varepsilon\approx10^{-3}$ can explain the discrepancy between theory and experiment
on the muon anomalous magnetic moment $\left(g-2\right)_{\mu}$ \cite{key-2}.


\section{The experiment}

Approved by the INFN at the end of 2015, PADME will search
for DP that decays into invisible channels (DM or long lived $A'$
independently of the decay products nature) \cite{key-3,key-4}. The experiment is designed
to detect as missing energy the $A'$ produced in the reaction $e^{+}\, e^{-}\rightarrow A'\,\gamma$,
where $e^{+}$, provided by the LNF linac, inping on a
fixed target. Since the initial state kinematic
is known ($\vec{P}_{e^{-}}$ and $\vec{P}_{beam}$ for the $e^{-}$
and the $e^{+}$, respectively), the $A'$ invariant mass can be determined measuring the
photon in the final state ($\vec{P}_{\gamma}$):

\[
M_{miss}^{2}=\left(\vec{P}_{e^{-}}+\vec{P}_{beam}-\vec{P}_{\gamma}\right)^{2},
\]

where $\vec{P}_{e^{-}}=\vec{0}$ and $\vec{P}_{beam}$ has the nominal
value of $550\,\mbox{MeV}$ along the z-axis.

\begin{figure}[!h]
\centering{}\includegraphics{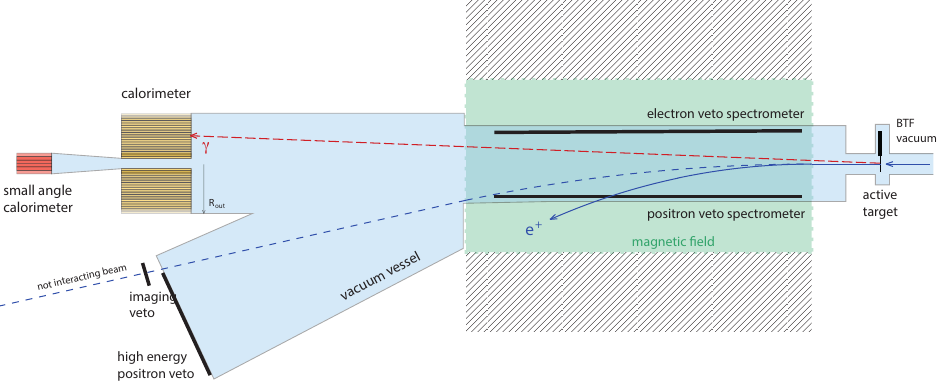}\caption{\label{fig:detector}PADME detector layout. Dimensions are $\mbox{L}=1\,\mbox{m}$,
$\mbox{D}=3\,\mbox{m}$ and $\mbox{R}_{\mbox{out}}=29\,\mbox{cm}$,
while the smal angle calorimeter is a square of $14\,\mbox{cm}$ side.}
\end{figure}

Figure \ref{fig:detector} shows the experiment layout.
The detector consists in an active target (diamond with graphitic
strips to determine the beam position), a MBP-S dipole, that deflects
the exaust beam and direct the $e^{+}$ that lost energy (the majority
for bremsstrahlung) towards the vetoes, an electromagnetic calorimeter
($616$ $2\times2\times22\,\mbox{cm}^{3}$ BGO with energetic
resolution $\sim\frac{(1-2)\%}{\sqrt{E}}$), that measures the
energy and the angle of produced $\gamma$, with a central opening to let bremsstrahlung
radiation to pass, which is then identified by the small angle
calorimeter ($49$ $2\times2\times20\,\mbox{cm}^{3}$
SF57 leaded glasses).

Main backgrounds to be reduced with the detector geometry 
are the $e^{+}\,e^{-}$ annihilation into $2$ or $3$ $\gamma$ and 
the bremsstrahlung, while the pile-up, being connected to the bunch density, 
will be the most important limiting factor to the PADME sensitivity.

\begin{figure}[h]
\centering{}\includegraphics[scale=0.4]{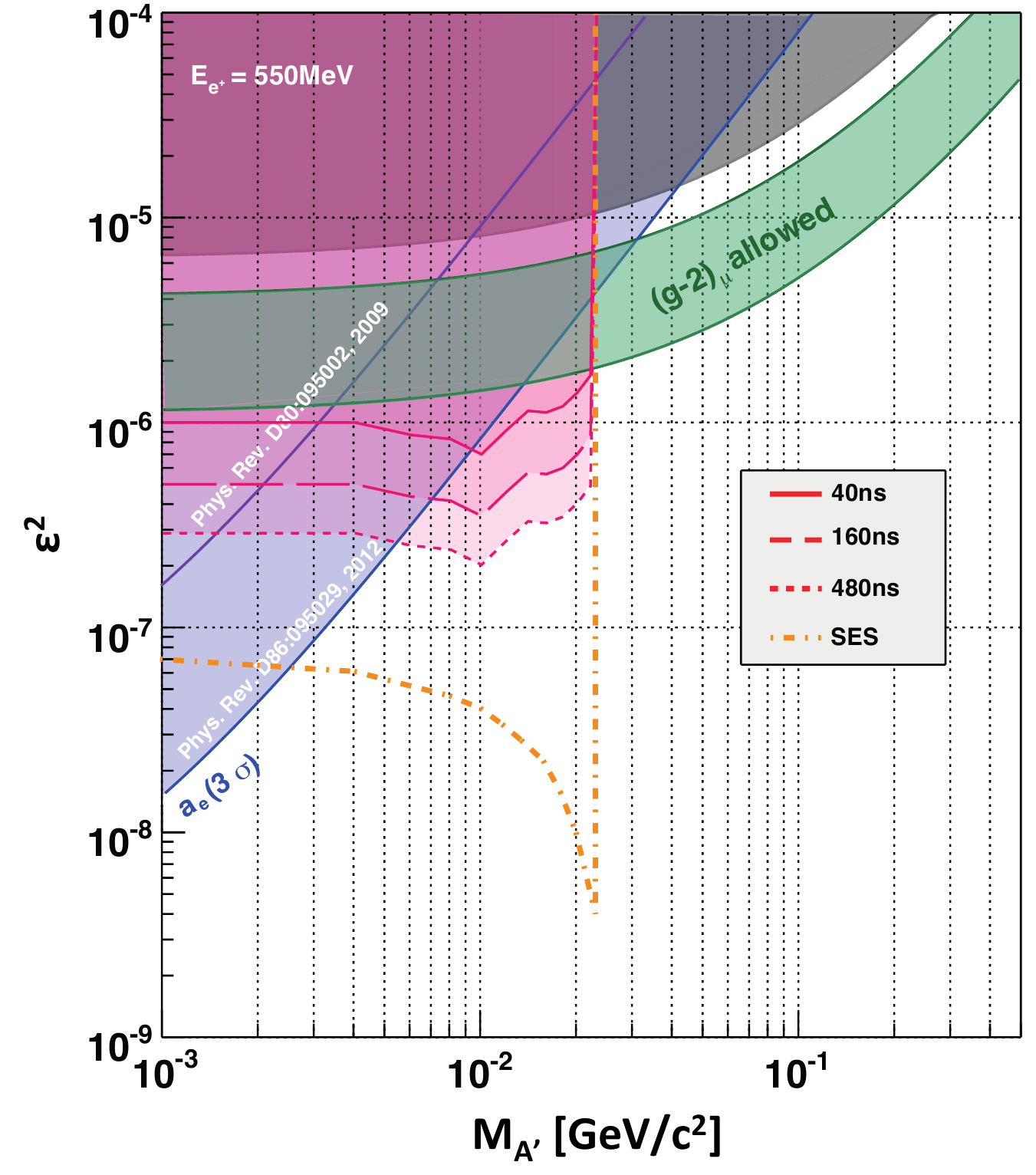}\caption{\label{fig:sensitivity}PADME sensitivity for different bunch
lengths and in absence of background (SES) at the nominal beam energy
of $550\,\mbox{MeV}$.}
\end{figure}

The collaboration aims to build the detector by the end of 2017
and to collect $10^{13}$ $550\,\mbox{MeV}$ positrons on target by
the end of 2018, reaching a sensitivity on the DP coupling constant
to the SM charge $\varepsilon$ of $\sim10^{-3}$ and a DP mass up to
$23.7\,\mbox{MeV}$ (see Fig. \ref{fig:sensitivity}).

\FloatBarrier

\end{document}